      [1999/12/01 v1.4c Il Nuovo Cimento]
\documentclass{cimento}

\usepackage{graphicx}
\usepackage{latexsym}

\title{Optical Camera with high temporal resolution to search for 
transients in the wide field}

\shorttitle{Optical Camera with high temporal resolution}

\author{
S.~Karpov\from{sao},
G.~Beskin\from{sao},
A.~Biryukov\from{sai},
S.~Bondar\from{ipp},
K.~Hurley\from{ssl},
E.~Ivanov\from{ipp},
E.~Katkova\from{ipp}, 
A.~Pozanenko\from{iki}
\atque
I.~Zolotukhin\from{sai},
}
\shortauthor{S.~Karpov et al}
\instlist{
\inst{sao}
SAO RAS, Nizhnij
Arkhyz, Karachai-Cherkessia, Russia, 369167
\inst{sai}
Sternberg Astronomical Institute of Moscow State University,
Universitetskij pr. 13, Moscow, Russia, 117415
\inst{ipp}
Research Institute for Precision Instrumentation, Moscow, Russia
\inst{ssl}
Space Science Laboratory, University of California, Berkeley, USA
\inst{iki}
Space Research Institute, Profsojuznaja 84/32, Moscow, Russia, 117997
}

\PACSes{
\PACSit{95.55.Cs}{Ground-based ultraviolet, optical and infrared telescopes}
\PACSit{98.70.Rz}{gamma-ray sources; gamma-ray bursts}
}

\begin{document}

\maketitle

\begin{abstract}
The wide field optical camera with high temporal resolution for the 
continuous monitoring of the sky in order to catch the initial 
stages of GRBs is described.
\end{abstract}


Gamma-ray bursts (GRBs) are one of the most powerful transient events
in the Universe which are probably related to the compact relativistic
objects. Fine time structure of GRB emission is defined by the 
properties of their central engine.

At the same time a number of models predict generation of considerable
optical flux synchronous with GRB event which can achieve $10-12^m$ for
0.2 sec \cite{eichler} or even $8-9^m$ for $0.1-10$ sec \cite{liang}.
In the model of early afterglow in the wind shell the optical flash is
expected to be as bright as $9-10^m$ in $0.2-0.5$ sec with $0.5-60$ 
sec lag from proper GRB event \cite{wu}.

Thus, search and study with high temporal resolution of optical
transients (OTs) accompanying GRBs, can provide statistically reliable
information about the nature of these phenomena.  To be successful such
observations have to be carried out independently of alerts receiving
from space borne gamma-ray telescopes and use optical instruments with
a wide field of view. As a side effect such observations also gives the
possibility to detect and investigate short stochastic  flares of
different variable objects -- SNs, flare stars, CVs, X-ray binaries  and
NEOs, natural and artificial. 

Simple analysis of the technical parameters of a camera for such a task
may be performed as follows.

It is well known that detection limit in the observations with CCD is
determine noise of the detector and the sky background.                

For wide-field instruments the last case is realized. In the V band, for the
atmospheric and optics transparencies of 0.5 and 0.7 correspondingly, and for
the sky background of 20 mag/$\Box^{''}$ the detection limit is
$S = 4.9\frac{aKl}{DF\sqrt{et}}$ photons/cm$^2$/s, or, in magnitudes,
$m = 13 - 2.5\lg\left[\frac{aKl}{DF\sqrt{et}}\right]$. We used flux calibration from
\cite{fukujita} and the following notation: $a$ - detecion threshold in 3$\sigma$,
$l$ - pixel size in 6.45 microns, $D$ - telescope diameter in 10 cm, $F$ - focal
length in 20 cm, $e$ - detector quantum efficiency in 0.6, $t$ - exposure time in
0.13 s. $K$ is the scaling coefficient in units of 4, which characterizes the
reduction of the focal image linear size by the special unit - the taper, or
the image intensifier, - to use the a single CCD to cover the whole original
telescope field of view.
For example, in our Camera the reduction of the FOV of 80 mm size by the image
intensifer and special objective by 7.6 times gives the possibility to
use standard 2/3 inch TV-CCD. Certainly, in any case the spatial resolution
of the main objective, reduction unit and the CCD have to be in agreement.

The rate $N$ of GRBs to be detected by the wide-field camera may be easily estimated
for standard slope of $-3/2$ of the ``$\log{N_{S}}-\log{S}$'' relation. The number of events
$N$ is then proportional to $N_{S} \propto S^{-3/2}$, where $S$ is the detection limit,
and the camera field of view $\Omega =  4.2\cdot10^4 P^2l^2K^2 / F^2$ arcsec$^2$, where
$P$ is the CCD size in pixels. So,
$N \propto N_{S}\Omega \propto S^{-3/2}P^2l^2K^2/F^2 \propto a^{-\frac32}e^{\frac34}t^{\frac34}A^{\frac12}K^{\frac12}DP^2l^{\frac12}$.

Here $A=D/F$ is the aperture of the system, which should be as high as
possible, but can hardly exceed 1 for the acceptable aberrations in the
wide field.

The important consequence of these formulae is that the GRB rate
depends on $D$ linearly and
quadratically on $P$ (thus, the field of view size). It means that for
higher GRB count rate the wider field of view is more important than
the larger objective diameter. Also it is clear that the wide field
may be achieved only by means of special methods of focal scale reduction
(tapers, image intensifiers) which enlarges $K$ (for the fixed maximal $A$).
Certainly, main objective has to give good quality FOV $\Omega$ corresponding to
the detector size $P$. That condition is much easier to achieve for smaller $D$.


The Camera has been created as a realization of ideas formulated above.
Its parameters are given in Table \ref{params}. The Camera has
$A=D/F=1/1.2=0.83$ and  uses the image intensifier to lower the focal scale
($K=1.9$, $e=0.17$), so the theoretical estimation of limiting magnitude for it is determined by the
sky background noise and is equal to 11.6$^{\rm{m}}$ for the 0.13 sec
exposure. The TV-CCD used (VS-CTT285-2001) is able to operate in 7.5 Hz
frame rate mode with such an exposure.

\begin{table}
\caption{Main parameters of the Camera. \label{params}}
\begin{narrowtabular}{0cm}{lr|lr|lr}
\hline
\multicolumn{2}{c|}{Main objective} & \multicolumn{2}{c|}{Intensifier} & \multicolumn{2}{c}{CCD}\\
\hline
Diameter &	150 mm		& Photocathode 		& S 25 & Dimensions & 1280x1024 pix \\
Focus &    	180 mm  	& Photocathode diameter & 90 mm	& Image scale & $56''$ / pix \\
D/F & 	   	1/1.2		& Gain 			& 150 	& Exposures  & 0.13-10 sec \\
Field of view & 21x16 deg 	& Scaling coefficient	& 7.6 / 1 & Pixel size & 6.45 micron \\
  &                       	& Quantum efficiency 	& 10\% 	& Data stream & 400 Gb / night \\
\hline
\end{narrowtabular}
\end{table}

\begin{figure}
\includegraphics[width=6.5cm]{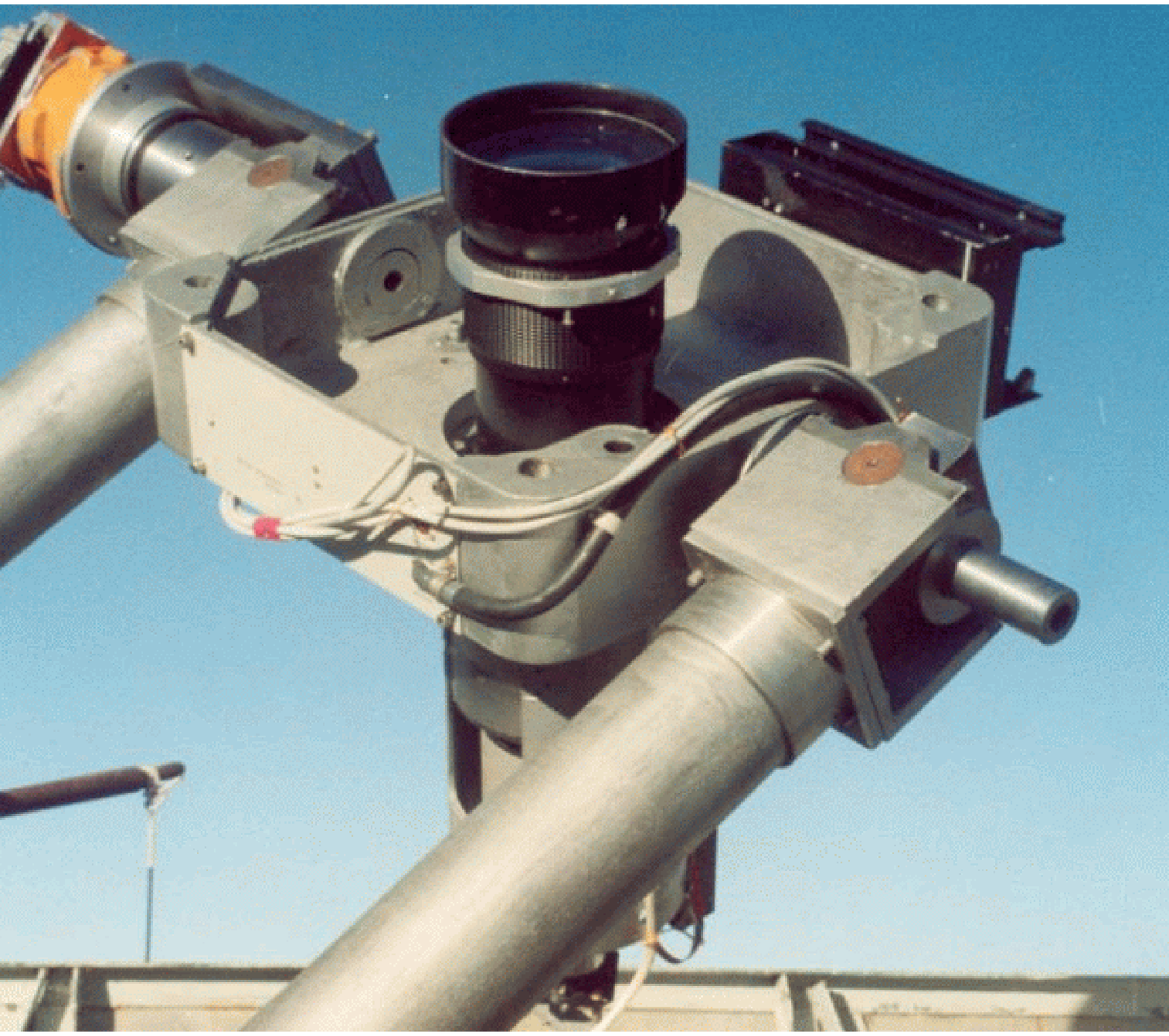}
\includegraphics[width=6.5cm]{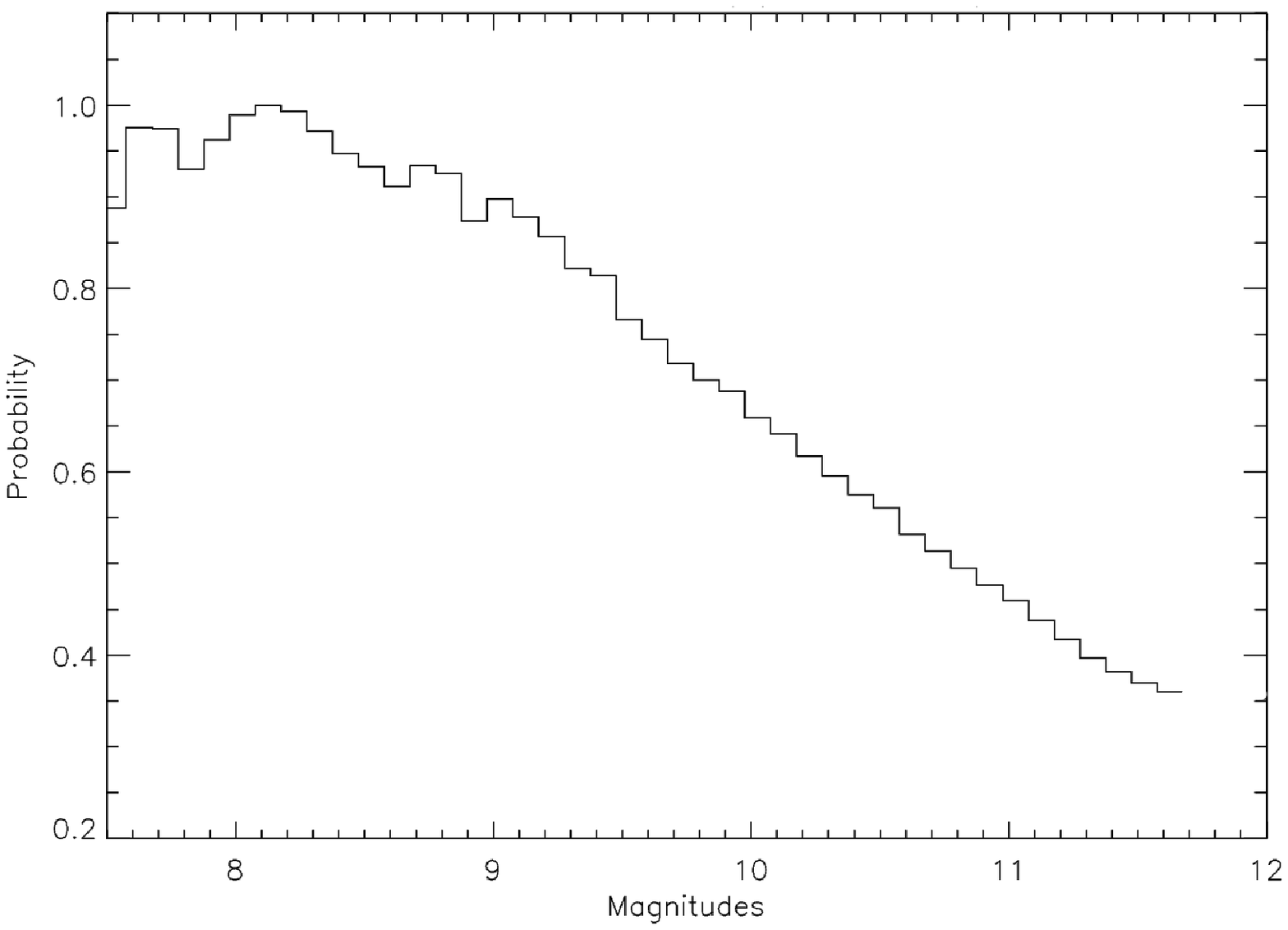}
\caption{
Left pane -- photo of a Camera installed in Northern Caucasus near the
6-m  telescope. 
Right pane -- typical detection efficiency (the probability for the
object  of given magnitude to be detected) for the objects in
dependence on magnitude. } 
\label{okh} 
\end{figure}

\begin{figure}
\includegraphics[width=6.5cm]{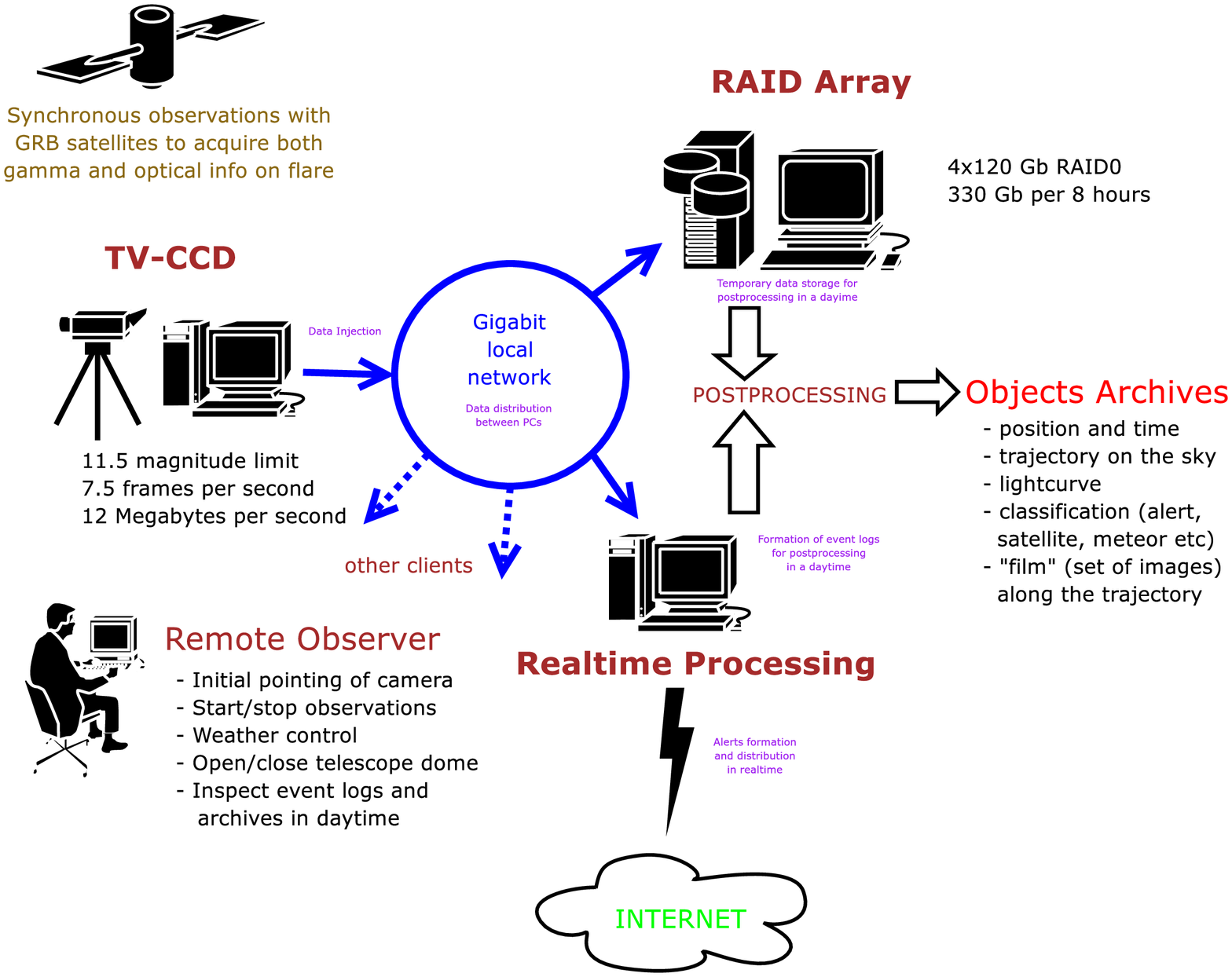}
\includegraphics[width=6.5cm]{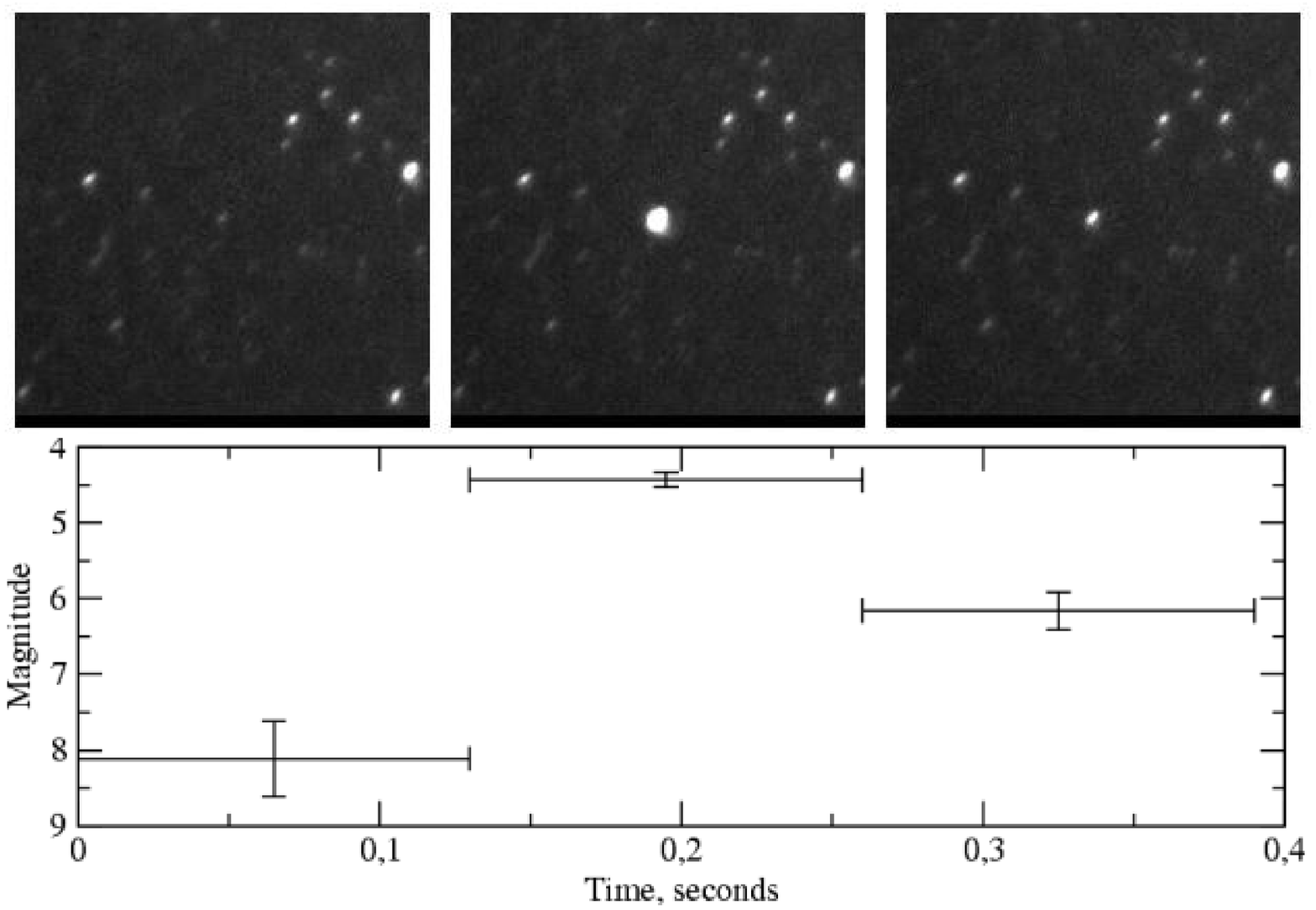}
\caption{
Left pane -- global scheme of the Camera hardware and software complex and 
data flow. Right pane -- example of a short flare detected by the camera. 
Total length of the event is 0.4 sec (seen on 3 successive frames).
}
\label{dataflow}
\end{figure}

  The observational data from the TV-CCD are transmitted (See Fig.
\ref{dataflow})  to the local PC which  brodcasts it through the LAN to
the storage -- RAID array with 480 Gb capacity --  and to the real-time
processing box. The data flow rate for the system is about 13 Mb/sec.
In order to process such a data stream in  real time no standard
reduction routines applied usually for field photometry  and source
extraction may be used. For this reason special software complex  for
detection and investigation of OTs has been created.

  The software is installed at the three PCs and operated by   WINDOWS
and LINUX OSes. Incoming information is a sample of 1280x1024 pixels
CCD frames with exposure time of 0.13 sec. The software performs  the
data reduction in real time -- detection and classification of OTs,
determination of their equatorial coordinates and magnitudes,  their
possible identification with known objects and transfer of information 
about OTs (alerts) to the local and global networks.

  The OT detection algorithm is based on the comparison of current
frame  with one averaged over 100 previous frames and is able to detect
and classify any transient that is seen on three successive frames
(i.e. with duration of  0.4 sec),  determine its shape, trajectory and
light curve and cross-correlate it with catalogues of known transient
objects such as stars and satellites. When transient doesn't match the
catalogues and doesn't look like a meteor the system is able to send
its information to robotic  telescopes and/or global networks.

The example of a short (0.4 sec) transient detected by the system is
shown on  Fig.\ref{dataflow}. The transient shown is due to
geostationary satellite.

As it has been noted above, the faintest detectable object has
11.5$^{\rm{m}}$  in a band close to V.  This limit may be increased up
to 12.5 - 14$^{\rm{m}}$ by  simultaneous analysis of sums of large
number of frames (10 - 100) by the cost of temporal  resolution lose.


Nearly all instruments for gamma-ray bursts related OTs (prompt
emission and afterglow) search currently in operation are divided into
two classes - the trigger based relatively large (25-100 cm) telescopes
with up to 10 deg$^2$ FOV and the monitoring cameras, consisting of 1-4
small objectives (5-10 cm) with 200-1500 deg$^2$ FOV. Both classes are
equipped with standart CCDs with exposures larger than 10 sec (in rare
cases -- ROTSE III -- 5 sec). The instruments of the first class need
30-60 sec for the pointing and thus can't begin the observations until
the  GRB fade off, but they are essential for the detection and study
of early afterglows (1-10 min after the trigger) due to ability to
perform accurate photometry of 17-20$^{\rm{m}}$ objects on the 5-60 sec
timescale \cite{rotse,tarot}. At the same time the monitoring systems
are able to detect 10-12$^{\rm{m}}$  OTs only for the 10-60
sec\cite{pi,raptor}. The Camera described has the 2$^{\rm{m}}$ better
sensitivity on the same time scale and even with 3-5 smaller FOV is
able to detect 3-5 times larger amount of OTs with similar duration
(for -3/2 distribution law), but also it is able to detect the flares
with 0.13-5 sec time scale, undetectable by the all other systems.


The Camera is placed in the Northern Caucasus (near the 6-m optical
telescope BTA)  at the height of 2030 m above the sea level. Since 2003
it monitors on the regular basis the part of HETE-2 gamma-ray satellite
field of view. For the whole  period of observations (approx. 150 good
nights) no GRB triggers hits the monitored  field. The Camera detects
the large number of meteors (approx. 300) and satellites (approx. 150)
each night. The work is in progress.

\acknowledgments

This work was supported by grants of CRDF (No. RP1-2394-MO-02), RFBR
(No. 04-02-17555), and INTAS (04-78-7366). S.K. thanks Russian Science Support Foundation for support.
G.B. thanks Cariplo Foundation for the scholarship
and  Merate Observatory for hospitality.

\end{document}